\documentclass[twocolumn,showpacs,prl,aps,superscriptaddress]{revtex4}

\usepackage{epsfig}
\usepackage{graphicx}
\usepackage{dcolumn}
\usepackage{bm} 
\newcommand{\beq}{\begin{equation}}
\newcommand{\eeq}{\end{equation}}
\newcommand{\beqa}{\begin{eqnarray}}
\newcommand{\eeqa}{\end{eqnarray}}

\newcommand{\om}{\omega}

\def\ol#1{{ Opt.\ Lett.} {\bf#1}}

\def\jpb#1{{ J.\ Phys.\ B} {\bf#1}}

\def\natphys#1{{ Nature\ Phys.\ } {\bf#1}}
\def\njp#1{{ New J.\ Phys.\ } {\bf#1}}
\def\pra#1{{ Phys.\ Rev. A\/} {\bf#1}}

\def\prl#1{{ Phys.\ Rev.\ Lett.} {\bf#1}}

\begin{document}

\title{Sequential Tunneling vs. Electron Correlation in Multiple Photo-ionization}

\author{X. Wang}
\affiliation{ Rochester Theory Center and Department of Physics \& Astronomy, University of Rochester, Rochester, New York 14627, USA}

\author{J. Tian}
\affiliation{ Rochester Theory Center and Department of Physics \& Astronomy, University of Rochester, Rochester, New York 14627, USA}

\author{A. N. Pfeiffer}
\affiliation{Chemical Sciences Division, Lawrence Berkeley National Laboratory, Berkeley, California 94720, USA}

\author{C. Cirelli}
\affiliation{Physics Department, ETH Zurich, CH-8093 Zurich, Switzerland}

\author{U. Keller}
\affiliation{Physics Department, ETH Zurich, CH-8093 Zurich, Switzerland}

\author{J.\ H.\ Eberly}
\affiliation{ Rochester Theory Center and Department of Physics \& Astronomy, University of Rochester, Rochester, New York 14627, USA}

\date{\today}

\begin{abstract} We take advantage of the information provided by use of elliptical polarization in a recent two-electron release time experiment [A.N. Pfeiffer {\it et al.}, Nature Physics {\bf 7}, 428 (2011)]. This allows a comparative test of the currently dominant conjectures regarding independent-electron tunneling theory vs. fully electron-correlated classical release theory to describe electron release in strong-field double ionization.
\end{abstract}


\maketitle


Atoms exposed to very strong laser fields are subject to ionization. With some experimental support \cite{Augst-etal}, quantum tunneling is generally conjectured as the physical process that is responsible for such ionization: an electron can be freed by ``tunneling through'' the potential barrier generated by the combination of the atomic Coulomb potential and the laser electric potential.

Theories have been proposed to predict the tunneling rate of an atom exposed to a laser field \cite{Keldysh, PPT, ADK}. These theories are based explicitly on an assumption that the laser electric field is weaker than the over-barrier field of the ionizing electron. The errors of these theories are difficult to estimate, but certainly grow substantially as the laser field becomes stronger, while the theories continue to be used for guidance. However, their applicability is now beginning to be open to quantitative test by comparison with recent experiments that record electron release times using some of the advantages provided by elliptical polarization of the laser field (for an overview, see \cite{Wang-EberlyPRA}).

To extend the applicability of the tunneling theories and at the same time keep the simplicity and convenience of the analytical form of the well-known Ammosov-Delone-Krainov tunneling formula \cite{ADK}, Tong and Lin \cite{Tong-Lin} proposed an augmented quantum tunneling (AQT) formula in which an additional exponential correction factor is added to account for the strong-field saturation effect. The correction factor in the AQT formula contains an empirical parameter $\alpha$ to be determined for each atomic and ionic species.

Under the usual tunneling conjecture, an electron's release time is not really open to question. Because of the exponential dependence of the emission rate on laser field strength predicted by tunneling theories, the electron is almost certainly released at the peak of the laser pulse. Modern laser technologies can now routinely generate laser fields that are strong enough to ionize multiple electrons from an atom. These laser fields can be near to or higher than the over-barrier field of the first electron or of the second electron, so an electron can be emitted earlier than the peak of the pulse, and determining multiple emission times becomes a viable route for testing.

\begin{figure} [t!]
  \includegraphics[width=4.2cm]{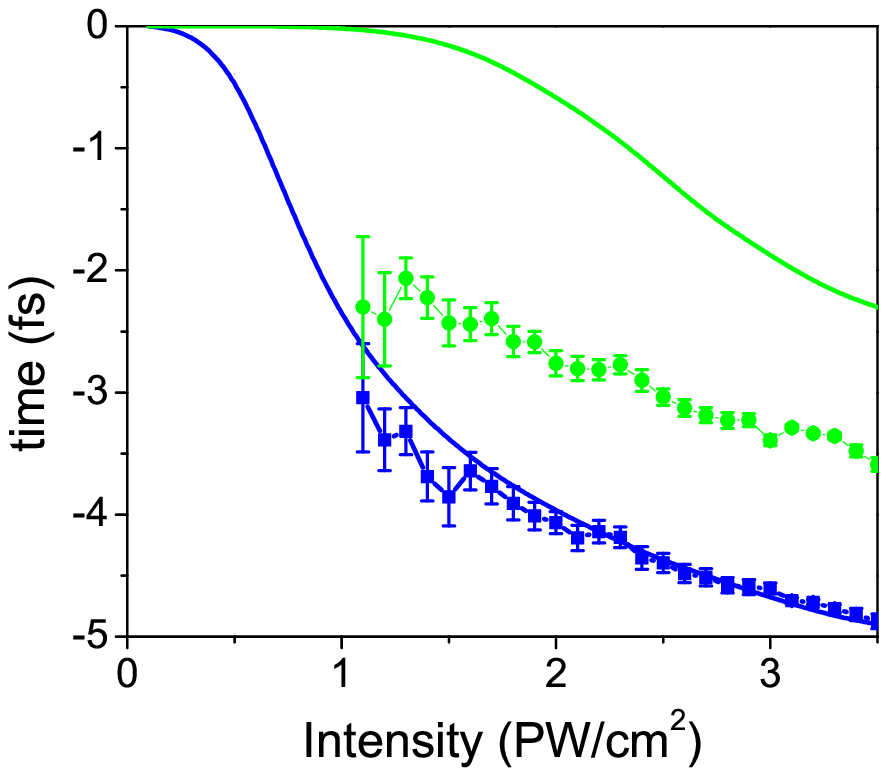}
  \includegraphics[width=4.2cm]{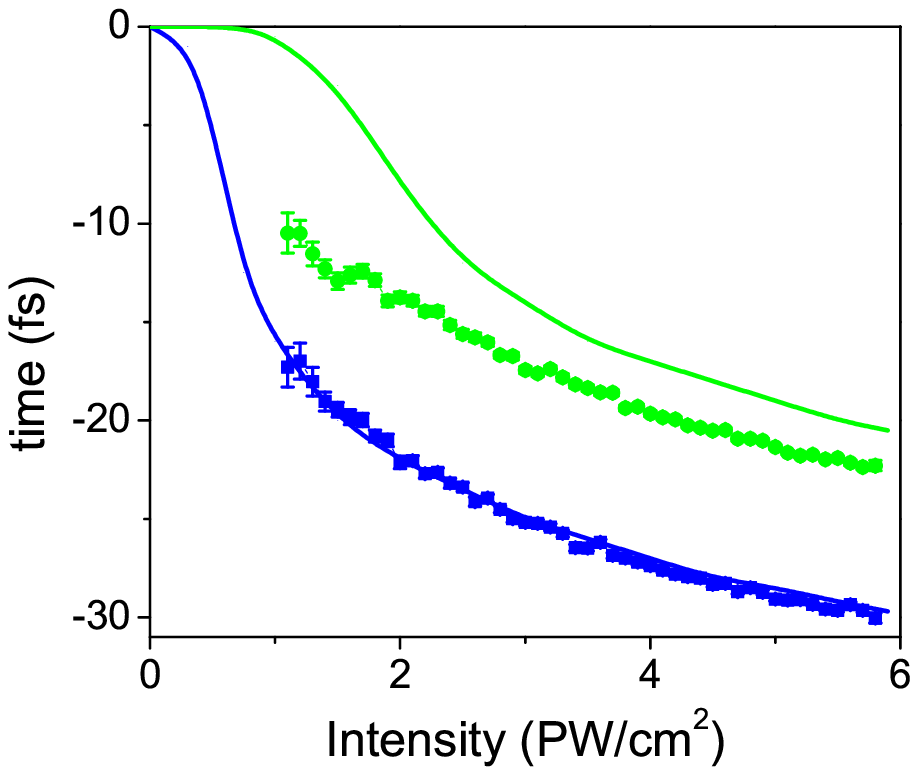}
  \caption{Comparison of the AQT predictions (solid lines) of emission times with experiment (dots with error bars). The left panel is for the 7fs pulse and the right panel is for the 33fs pulse. Blue color is for first ionization and green color is for second ionization. Adapted from \cite{Pfeiffer-etalNatPh}.  }\label{f.t1t2}
\end{figure}

The AQT formula has been used by Pfeiffer {\it et al.} \cite{Pfeiffer-etalNatPh} in connection with their measurements of the sequential release times in double ionization of Ar, using elliptical polarization. To deal with two-electron ionization, the AQT theory uses a single-active-electron (SAE) approximation, which assumes that only one electron is active at each time, with other electrons playing no role but screening the nucleus. Only after the first electron is ionized does a second electron become active. Comparison of the AQT predictions with experiments show good agreement with the observed release time of the first electron, while they predict substantially later times than observed for the second electron. This is clearly seen in Fig. \ref{f.t1t2}. It has therefore been conjectured that the SAE (or independent-electron) approximation breaks down, meaning that electron-electron correlations play a role. However the specific role that they play has not been explained.

The same conjecture of required e-e correlation has also been advanced for another experiment, with a similar reasoning \cite{Pfeiffer-etalNJP}. In that case, the tunneling theory fails to explain an oscillation feature of the measured ratio between parallel and anti-parallel sequential double ionization (SDI) emissions (i.e., between SDI with the two electrons released in the same or opposite directions). Interestingly, a fully correlated classical ensemble simulation based on solution of the relevant time-dependent Newton equations (TDNE) qualitatively reproduces the oscillation feature \cite{Wang-EberlyarX}. The physical role of the e-e correlations was again not clearly identified.

Moreover, Zhou {\it et al.} \cite{Zhou-etal} recently used a classical ensemble simulation, again employing TDNE solutions of the relevant equations, and reported the first successful reproduction of the measured electron release times. They modelled each atom in their ensemble as containing two fully correlated electrons able to interact steadily with each other, as well as with the laser field and the ion. But the specific role of the e-e correlations leading to the experimental agreement were not identified or analyzed.

From this short review, one sees that conjectures have been advanced that e-e correlations must necessarily be present on either of two different grounds: because (1) the independent-electron SAE-tunneling AQT approach disagrees with the experiments, and/or (2) the fully-correlated classical simulation agrees with the experiments.

Of course, making explanatory conjectures such as the need for e-e correlation should be encouraged. But the lack of an understanding of the role for such correlations substantially weakens the conjectures. There is a reason for this. As we report here, systematic calculations show, surprisingly, that the need for e-e correlation on the one hand, or quantum tunneling on the other hand, can both be discarded in treating electron release timing. Without either one, quantitative agreement with experimental release times can still be achieved.

Classical ensemble calculations (traditionally fully-correlated) have previously been described in detail \cite{Panfili-etal, Su-Eberly}. Recently the classical ensemble method has been extended to include elliptical polarization \cite{Wang-EberlyPRL09, Wang-EberlyPRL10, Wang-EberlyNJP10} and good agreement with experiment has been achieved \cite{Maharjan-etal, Pfeiffer-etalNatPh}. In the calculations reported here, the same method is implemented, however while enforcing SAE evolution. This is done by numerically permitting only one electron to be actively involved in the ionization process, and then after the first electron is removed by the laser field a second electron is allowed to become active and subject to removal from the residual ion. This process can be extended to additional electrons as desired.

\begin{figure} [t!]
\includegraphics[width=4.2cm]{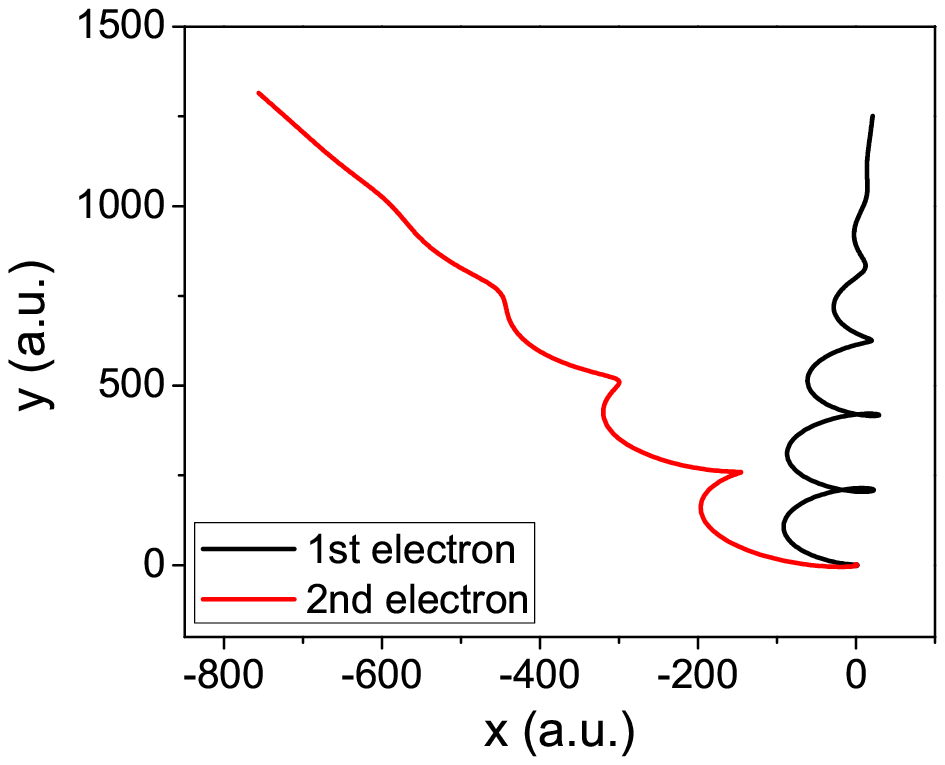}
\includegraphics[width=4.2cm]{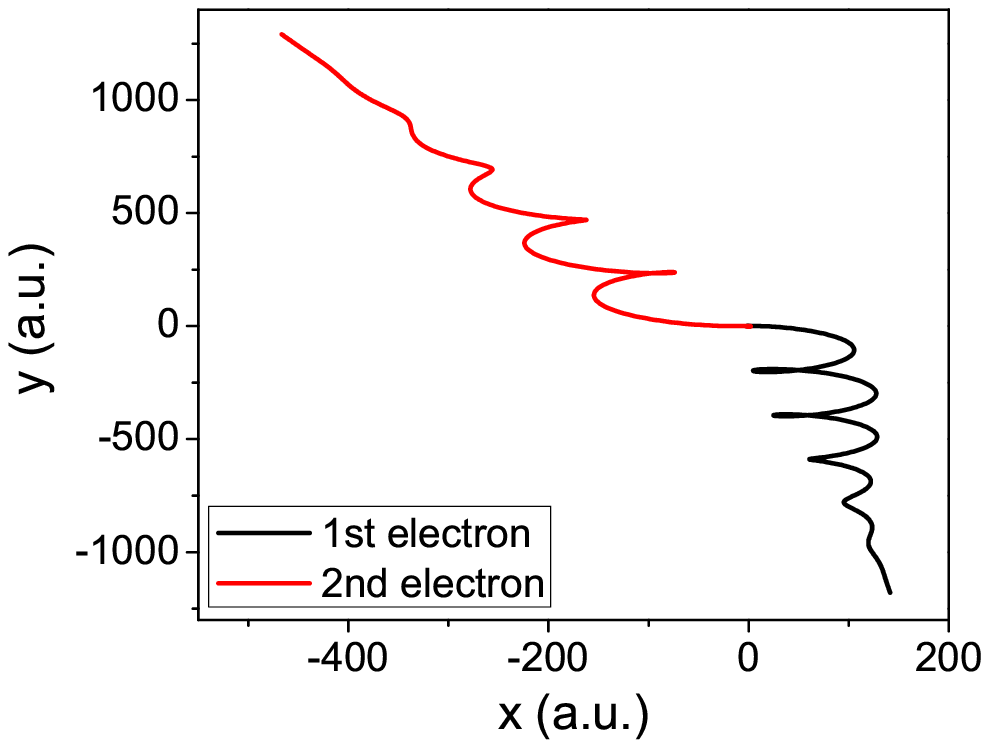}
\caption{Two typical SDI trajectories with elliptical polarization. The left panel shows a time delay of an integer number of cycles between the two emissions, and the right panel shows a time delay of an odd multiple of half cycles between the two emissions.} \label{f.trajs}
\end{figure}

Without difficulty, the Newtonian equations of motion can be integrated numerically under the SAE constraint and the positions and the momenta of the two electrons are recorded at each time step until the end of the pulse. At the end of the pulse, ionization trajectories with both electrons at least 10 a.u. away from the ion core can be collected and further analyzed. Typical two-electron release trajectories are shown in Fig. \ref{f.trajs}, corresponding to a time delay of an integer number of cycles between the two emissions (left) and of an odd multiple of half cycles between the two emissions (right). Trajectories as shown in the left and right panels usually lead, respectively, to larger and smaller net ion momenta along the minor direction, analogous to the non-Z and Z trajectories found with linear polarization \cite{Ho-etalPRL05}. Detailed explanations can be found in \cite{Wang-EberlyPRL09}.

Some specific details of the classical SAE simulation procedure are described as follows. Initially an ensemble of classically modeled atoms with only one electron is generated. The energy of this electron is set to the negative of the first ionization potential (noted as $IP_1$)
\beq
    E_1 = \frac{\mathbf{p}_1^2}{2} - \frac{1}{\sqrt{\mathbf{r}_1^2+1}} \equiv -IP_1,
\eeq
where $\mathbf{p}_1$ and $\mathbf{r}_1$ are the momentum and the position of the electron. Note that the ion core Coulomb potential has been softened to avoid numerical singularities \cite{Su-Eberly}. Argon, which is the target atom used in the Pfeiffer experiment, has $IP_1 = 0.58$ a.u. Given the energy, then the position and the momentum of the electron are randomly assigned. Therefore the initial ensemble is a collection of model atoms with the same energy but different position and momentum configurations.

Then a laser pulse is turned on. We use the same laser parameters as in \cite{Pfeiffer-etalNatPh}. The laser field is elliptically polarized with ellipticity value $\varepsilon=$ 0.77. The pulse envelope has a Gaussian shape of duration (FWHM) 7fs or 33fs. The laser field has the form
\beq
    {\bf E}(t) = \frac{f(t)}{\sqrt{1+\varepsilon^2}} \left[ {\bf \widehat{e}_x} \sin(\om t + \varphi_0 ) + {\bf \widehat{e}_y} \varepsilon \cos(\om t + \varphi_0) \right],
\eeq
where $f(t) = E_0 \exp \left( -t^2/(2 \sigma^2) \right)$ is the pulse envelope. $E_0$, $\om$, and $\varepsilon$ are the amplitude, angular frequency, and ellipticity of the field. The carrier envelope phase is noted as $\varphi_0$, which is not stabilized and varies from pulse to pulse. Consistent with Refs. \cite{Pfeiffer-etalNJP,Wang-EberlyPRL09,Wang-EberlyNJP10}, the x direction has been selected as the major polarization direction and the y direction as the minor polarization direction.

The Hamiltonian of this electron is
\beq
    H=H(t)=\frac{\mathbf{p}_1^2}{2} - \frac{1}{\sqrt{\mathbf{r}_1^2+1}}+{\bf r}_1 \cdot {\bf E}(t).
\eeq
If the laser field is strong enough, the electron can be ionized and driven away from the vicinity of the ion core. Only then, in the SAE approximation, is a second electron allowed to be active \cite{Haan}. In our approach, a second electron is ``created" near the ion core when the first electron reaches a distance of 10 a.u.  A slight different choice of this distance will not affect our discussions at all.

The initialization of the second electron is similar to that of the first electron, and the energy of the second electron is
\beq
    E_2 = \frac{\mathbf{p}_2^2}{2} - \frac{2}{\sqrt{\mathbf{r}_2^2+1}} \equiv -IP_2,
\eeq
where $IP_2 = 1.02$ a.u. for the second electron. Given the energy value, the position and the momentum of this electron are randomly assigned.

\begin{figure} [t!]
  \includegraphics[width=4.2cm]{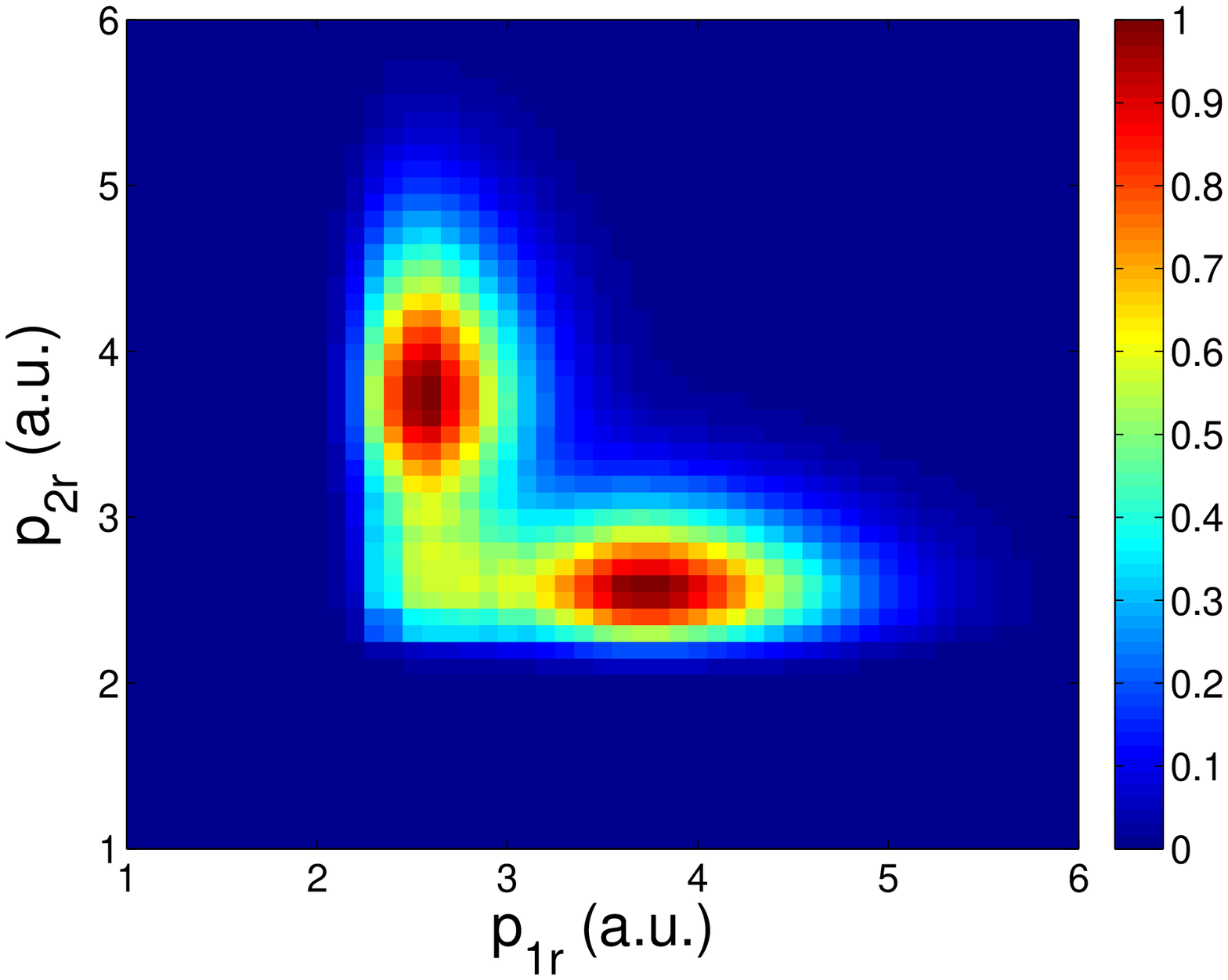}
  \includegraphics[width=4.2cm]{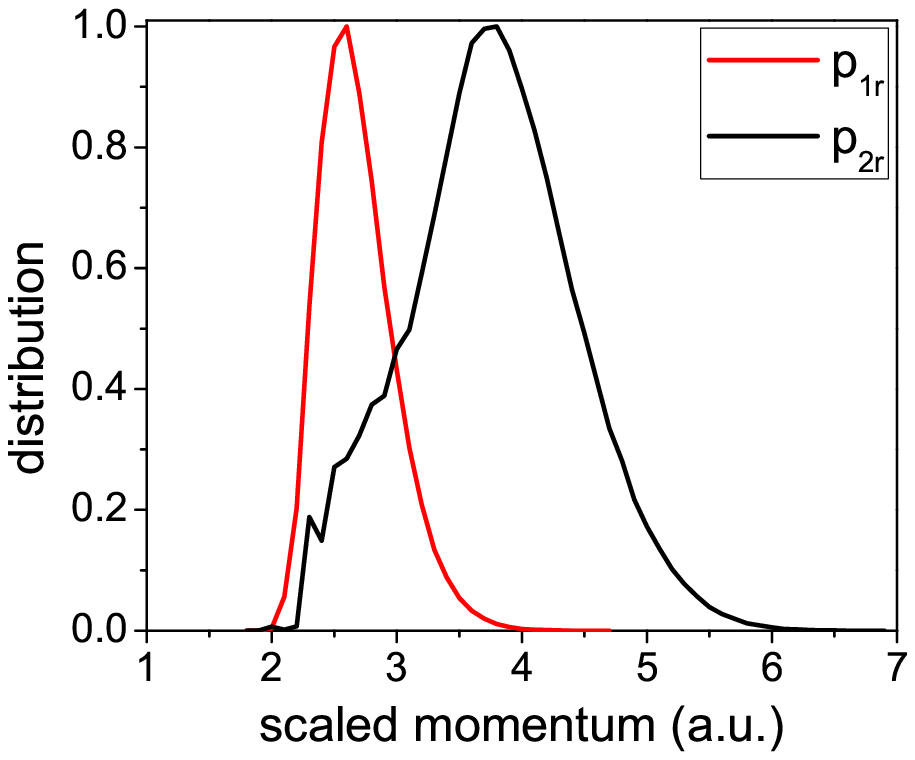}
  \caption{Left: a correlation spectrum of $p_{1r}$ and $p_{2r}$ for a 33fs pulse with peak intensity 4.5 PW/cm$^2$. Right: reduced distributions of $p_{1r}$ and $p_{2r}$, under the same condition. Heights of peaks have been normalized to unity.}\label{f.p1rp2r}
\end{figure}

The second electron reacts immediately to all the forces on it. These forces include the Coulomb attraction from the ion core, the laser electric force, and the repulsive force from the first electron, which is already negligibly weak. The motion of the second electron is also governed by the classical TDNE. If the laser field is strong enough, the second electron can be ionized.

The momenta of the two ionized electrons at the end of the pulse are recorded as \{$p_{1x}$,$p_{1y}$,$p_{1z}$\} and \{$p_{2x}$,$p_{2y}$,$p_{2z}$\}. Two new scaled momenta are defined \cite{Pfeiffer-etalNatPh}:
\beqa
    p_{1r} &=& \sqrt{(1+\varepsilon^2)\left(p_{1x}^2/\varepsilon^2 + p_{1y}^2\right)}; \\
    p_{2r} &=& \sqrt{(1+\varepsilon^2)\left(p_{2x}^2/\varepsilon^2 + p_{2y}^2\right)}.
\eeqa
A correlated spectrum of $p_{1r}$ and $p_{2r}$ is shown in the left panel of Fig. \ref{f.p1rp2r} for the 33fs pulse with peak intensity 4.5 PW/cm$^2$. It is diagonally symmetric due to the fact that no information about the ionization order is pre-known experimentally. Nevertheless, because the first ionized electron has a smaller scaled momentum $p_r$ than the second ionized electron, the diagonally symmetric correlation spectrum  can be separated into individual distributions of the first and second ionized electrons, as shown in the right panel of Fig. \ref{f.p1rp2r}. The peak positions of the two distributions will be used to calculate the two emission times.

\begin{figure} [b!]
  \includegraphics[width=4.2cm]{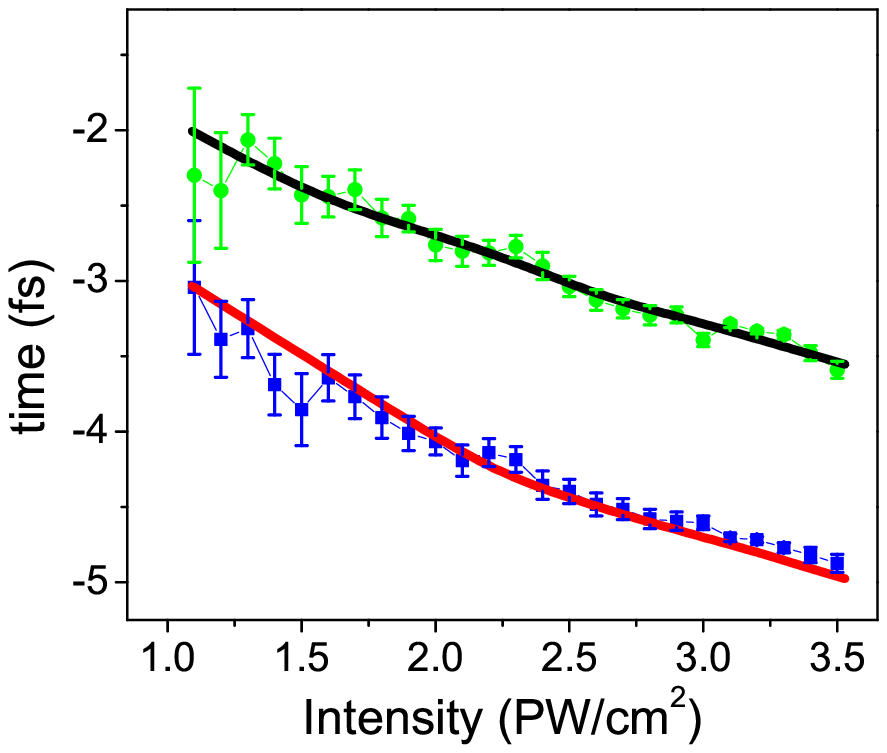}
  \includegraphics[width=4.2cm]{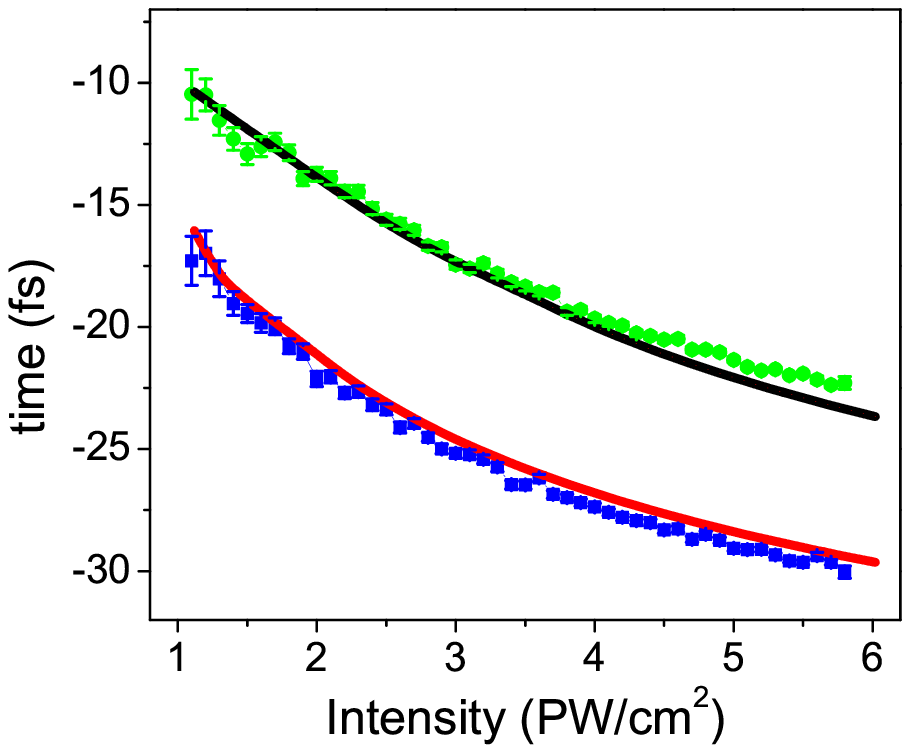}
  \caption{Comparison of TDNE predictions of emission times with experiment, obtained with new calculations, as explained in the text. The left panel is for the 7fs pulse and the right panel is for the 33fs pulse. Dots with error bars are the experimental data adapted from \cite{Pfeiffer-etalNatPh}. Lower and upper solid lines are the classical SAE predictions of the 1st and the 2nd ionization, respectively.}\label{f.t1t2_CL}
\end{figure}

Given that both ionizations happen before the peak of the pulse, the two emission times can be uniquely determined from the peak positions of the distributions of $p_{1r}$ and $p_{2r}$ (treated as Gaussian) \cite{Smolarski-etal}:
\beqa
    t_i = -\sigma \sqrt{2 \ln \left( \frac{E_0}{\om p_{ir}} \right)}, \quad \text{for } i=1,2. \label{e.pr2t}
\eeqa
The results of the calculated ionization times as a function of intensity are shown in Fig. \ref{f.t1t2_CL}, for both the 7fs and the 33fs pulses. Focal volume effects have been taken into account \cite{FocalVolume}, as done in \cite{Pfeiffer-etalNatPh} and \cite{Zhou-etal}. Quantitative agreement is highly satisfactory over the full experimental intensity range.

The physical reason for the ability of the classical simulation to match experimental results, as shown in Fig. \ref{f.t1t2_CL}, while the AQT theory cannot, as shown in Fig. \ref{f.t1t2}, remains an important open question. The fact is that both are based on the SAE approximation, so their key difference is simply rejection or adoption of initiation by tunneling. The other interesting open issue is in the agreement of our results not only with the experiments, but also with the calculations of Zhou {\it et al}. Those fully include e-e correlation, which our results suggest are surprisingly unimportant.

We need to emphasize that the above argument by no means proves that the classical model is correct while the AQT theory is wrong. One may equally argue that the AQT theory may be correct, while the disagreement between the Pfeiffer experiment and the AQT prediction may originate from e-e correlations or multielectron effects excluded from the AQT theory. We cannot deny this possibility but we do not like it. First, one always prefers simpler theories without many complications. And second, the conjectured e-e correlations or multielectron effects are quite difficult to identify, particularly as the results of the fully correlated theory of Zhou {\it et al}. is entirely in accord with our completely uncorrelated results.

To summarize, in our calculations we have discarded any remnant of either of the conjectures (1) or (2) identified at the outset. This was done quite simply. We adopted the usual classical ensemble theory, in order to strictly preclude tunneling. And then we numerically prevented e-e correlation. In this way we simultaneously tested both the importance of tunneling and the necessity of electron correlation. Classical calculations have regularly provided useful trajectory-based insights into strong-field ionization dynamics \cite{Ho-etalPRL05, Ho-Eberly06, Haan-etalPRL06, Haan-etalPRL08, MCU09, MCU10}, and we have no reason to suspect anything different here.

In conclusion, we have shown that the Pfeiffer {\it et al.} experiment on electron release times of SDI can be quantitatively reproduced by a tunneling-free, correlation-free classical analysis. The physical reason for the difference between the two SAE theories, our classical simulation and the AQT result, is yet an open question to answer.\\

Acknowledgement: This work was supported by DOE Grant No. DE-FG02-05ER15713, the NCCR Quantum Photonics (NCCR QP) and Molecular Ultrafast Science and Technology (NCCR MUST), research
instruments of the Swiss National Science Foundation (SNSF), ETH Research Grant ETH-03 09-2, and by the SNSF R'Equip grant 206021\_128551/1. A.N.P. is supported by the Laboratory Directed Research and Development Program at Lawrence Berkeley National Laboratory. X.W. acknowledges helpful discussions with Dr. Yueming Zhou.


\begin{thebibliography}{99}

\bibitem{Augst-etal} For example, see the early report: S. Augst, D. Strickland, D.D. Meyerhofer, S.L. Chin, and J.H. Eberly, \prl{63}, 2212 (1989).

\bibitem{Keldysh} L.V. Keldysh, Sov. Phys. JETP {\bf 20}, 1307 (1965).

\bibitem{PPT} A.M. Perelomov, V.S. Popov, and M.V. Terent'ev, Sov. Phys. JETP {\bf 23}, 924 (1966).

\bibitem{ADK} M.V. Ammosov, N.B. Delone, and V.P. Krainov, Sov. Phys. JETP {\bf 64}, 1191 (1986).

\bibitem{Wang-EberlyPRA} X. Wang and J.H. Eberly, \pra{86}, 013421 (2012).

\bibitem{Tong-Lin} X.M. Tong and C.D. Lin, \jpb{38}, 2593 (2005).

\bibitem{Pfeiffer-etalNatPh} A.N. Pfeiffer, C. Cirelli, M. Smolarski, R. D\"orner, and U. Keller, \natphys {7}, 428 (2011). See also commentary in K. Ueda and K. Ishikawa, \natphys {7}, 371 (2011).

\bibitem{Pfeiffer-etalNJP} A.N. Pfeiffer, C. Cirelli, M. Smolarski, X. Wang, J.H. Eberly, R. D\"orner, and U Keller, \njp{13}, 093008 (2011).

\bibitem{Wang-EberlyarX} X. Wang and J.H. Eberly, arXiv:1102.0221 (2011).

\bibitem{Zhou-etal} Y. Zhou, C. Huang, Q. Liao, and P. Lu, \prl{109}, 053004 (2012).

\bibitem{Panfili-etal} R. Panfili, J. H. Eberly, and S. L. Haan, Opt. Express {\bf 8}, 431 (2001).

\bibitem{Su-Eberly} See J. Javanainen, J. H. Eberly and Q. Su, \pra {38}, 3430 (1988) and Q. Su and J. H. Eberly, \pra {44}, 5997 (1991).

\bibitem{Wang-EberlyPRL09} X. Wang and J.H. Eberly, \prl{103}, 103007 (2009).

\bibitem{Wang-EberlyPRL10} X. Wang and J.H. Eberly, \prl{105}, 083001 (2010).

\bibitem{Wang-EberlyNJP10} X. Wang and J. H. Eberly, \njp{12}, 093047 (2010).

\bibitem{Maharjan-etal} C.M. Maharjan, A.S. Alnaser, X.M. Tong, B. Ulrich, P. Ranitovic, S. Ghimire, Z. Chang, I.V. Litvinyuk, and C.L. Cocke, \pra{72}, 041403(R) (2005).

\bibitem{Ho-etalPRL05} Phay J. Ho, R. Panfili, S.L. Haan, and J.H. Eberly, \prl{94}, 093002 (2005).

\bibitem{Haan} This process is similar to one employed in \cite{Haan-etalPRL08},  in which the model potential was numerically manipulated to test recollision scenarios.

\bibitem{Smolarski-etal} M. Smolarski, P. Eckle, U. Keller, and R. D\"orner, Opt. Express {\bf 18}, 17640 (2010).

\bibitem{FocalVolume} P. Wang, A. M. Sayler, K.D. Carnes, B.D. Esry, and I. Ben-Itzhak, \ol {30}, 664 (2005).

\bibitem{Ho-Eberly06} Phay J. Ho and J.H. Eberly, \prl {97}, 083001 (2006).

\bibitem{Haan-etalPRL06} S.L. Haan, L. Breen, A. Karim, and J.H. Eberly, \prl{97}, 103008 (2006).

\bibitem{Haan-etalPRL08} S.L. Haan, J.S. Van Dyke and Z.S. Smith, \prl{101}, 113001 (2008).

\bibitem{MCU09} F. Mauger, C. Chandre and T. Uzer, \prl{102}, 173002 (2009).

\bibitem{MCU10} F. Mauger, C. Chandre and T. Uzer, \prl{104}, 043005 (2010).

\end{thebibliography}
\end{document}